# Experiencing Utopia. A Positive Approach to Design Fiction.

**Judith Dörrenbächer**
Ubiquitous Design
University of Siegen
Siegen, Germany
judith.doerrenbaecher@uni-siegen.de

**Matthias Laschke**
Ubiquitous Design
University of Siegen
Siegen, Germany
matthias.laschke@uni-siegen.de

**Diana Löffler**
Ubiquitous Design
University of Siegen
Siegen, Germany
diana.loeffler@uni-siegen.de

**Ronda Ringfort**
Ubiquitous Design
University of Siegen
Siegen, Germany
ronda.ringfort@uni-siegen.de

**Sabrina Großkopp**
Ubiquitous Design
University of Siegen
Siegen, Germany
sabrina.grosskopp@uni-siegen.de

**Marc Hassenzahl**
Ubiquitous Design
University of Siegen
Siegen, Germany
marc.hassenzahl@uni-siegen.de

## Abstract
Design Fiction is known for its provocative and often dystopian speculations about the future. In this paper, we present an alternative approach that focuses primarily on the positive. We propose to imagine, enact, and evaluate utopia with participants. By doing so, we react to four main critiques concerning Design Fiction: (1) its negativity, (2) its contextlessness, (3) its elitist authorship, and (4) its missing evaluation methods.

## Author Keywords
Positive Design Fiction; utopia; enactment; performative methods.

## Critical Aspects of Design Fiction
Although the term *Design Fiction* initially described a literary method [24], today, it most often points to the widely popular practice of prototyping and visualizing artifacts of fictional worlds [5]. Most Design Fictions draw on the ideas of Anthony Dunne's and Fiona Raby's Critical Design [10,11] and Speculative Design [12]. These designs do neither optimize products nor solve problems but allow to reflect on a possible future [3,5,19]. To provoke discussions designers of Design Fictions often build on black humor, irony, and parody [6]. This tendency of being ironic, negative and fear-mongering has been an essential critique of Design

Fiction. For example, Tonkinwise states: "Everything they [Dunne, Raby & colleagues] make real is concerning at best and often just horrifying" ([25], p.187). He calls for fictions that are more suggestive: "[T]here should be much more readily identifiable moments of non-ironic endorsement, elements that make clear cases for what would be valuable (and not just sexy or fun) about these futures for significant sections of the population" ([25], p.186). And indeed, there are only a few Design Fiction approaches that explicitly reject irony to be destructive and argue for, e.g., a constructive critique [7], positive fictions [8], humorous, or 'seriously silly' design [6] instead.

A second critique of especially earlier attempts to Design Fiction [3,5,19] is that often only fictional artifacts are materialized, but not scenarios. The complex world of the artifact remains just imagination and is therefore hard to negotiate. This often results in vague anxieties. Just recently, some authors speak up for Design Fiction methods that embrace ambiguity, different perspectives, and complexity of fictional worlds [13,21,22].

A third critical issue with Design Fictions is that they are often presented in art galleries, magazines, or books [17], turning potential users and creators into passive spectators. Moreover, ongoing discussions accuse Design Fiction and the related Critical Design as being too elitist [16]. Designers often appear to act as moral agents who try to wise up people with their warnings inscribed into fictional renderings, 3D models, or photographs [4]. Consequently, a participatory version of Design Fiction is increasingly demanded and explored [1,9,13,14,17]. However, in most of those cases, participants do not create but only evaluate Design Fictions. Only a few approaches actually co-create Design Fiction by, e.g., creative writing [2], co-sketching [20] or enactments [9,13,26].

A fourth critique is that not many defined evaluation methods of Design Fiction exist. One of the few exceptions is called *Anticipatory Ethnography*, where experts (design ethnographers) observe fiction as if it were reality [18]. Beyond that, it is increasingly common to confront laypeople with Design Fictions and interview them afterward [1]. In all of those approaches Design Fiction is evaluated from a distanced point of view. People are observers of an extrinsic fiction, but they are not part of it.

In this position paper, we present an approach responding to our four described main critiques concerning Design Fiction: (1) its predominant negativity, (2) its contextless presentation, (3) its elitist authorship, and (4) its missing evaluation methods.

## A Co-Creation of Utopia

We describe our three-step approach (imagining utopia, calling utopia into being, evaluating utopia) in the following and how we applied it in a workshop on the topic of 'sustainability'. Five participants (3 male) took part.

*Step 1: Imagining Utopia instead of Dystopia*
We believe utopian thinking is essential to allow positive changes in society. Instead of provoking others with dystopian fictions, we need at least ideas of perfected versions of society. With our procedure, we build on theories such as Positive Psychology [23], Design for Wellbeing [14], and the already existing positive approaches to Design Fiction [6–8]. In the first step (Figure 1), we made participants imagine themselves to live in an utopian sustainable setting. We asked them to complete the following sentence: "In a sustainable society I feel positive, because […]". The participants came up with about 20 reasons, such as "[…] I can help out others with the resources I own." or "[…] I spend more time with people than with things." In a discussion we specified the positive emotions and realized that the participants imagined themselves to feel more related or popular in a sustainable utopia. Only after specifying their emotions, we discussed what kind of technology could allow them to experience such an utopia. One participant, for example, came up with the concept *Hyperpipe*, an infrastructure connecting all future households, allowing them to share physical goods.

*Step 2: Calling Utopia into Being with contextualized Enactments instead of watching fiction passively*
The term utopia, coming from Greek, means 'no place' or 'placeless place'. Utopia usually only exists in mind. We explore, what happens when utopia is materialized so that one is able to interact in it. We draw on the already mentioned approaches about tangible fiction [13,21,22] and participation in Design Fiction [2,15,16,20]. Thus, in the second step (Figure 2) of our approach we made the participants use some simple probes to co-construct utopia. We set some roles, elaborated a specific context and enacted the fictional world. For example, one participant acted as a salesperson of *Hyperpipe*, convincing a mayor of a small town to invest in it. Subsequently, a couple used *Hyperpipe* and sent winter clothes to people asking for it. While acting out utopia, several conflicts came up. For example, the couple started a conflict about what products they wanted to share. Sometimes we had to reset or change the initial utopian idea to make everybody (almost) happy again.

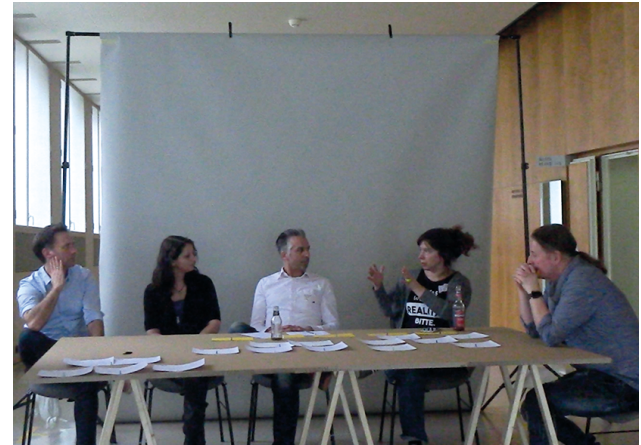
Figure 1: Participants, discussing positive emotions and possible technology of utopia.

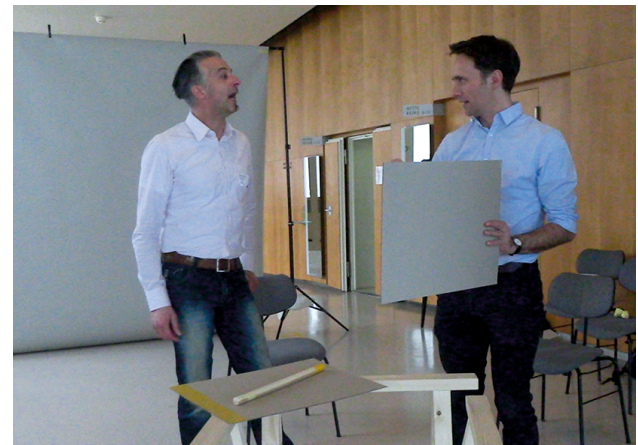
Figure 2: Citizens of utopian *Smalltown* using the sharing system *Hyperpipe*.

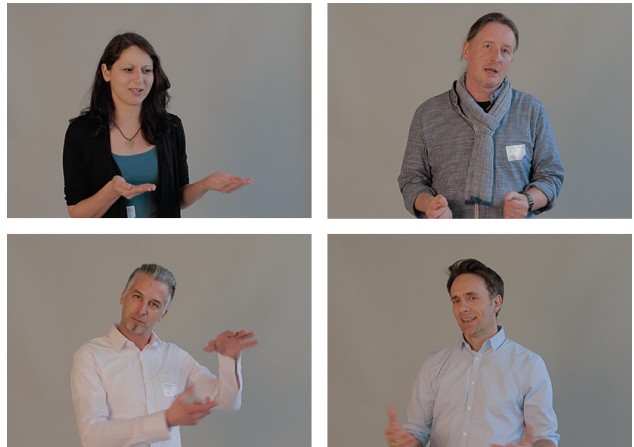

Figure 3: Participants evaluating the fictional technology from the view of their fictional character.

*Step 3: Evaluating Utopia from within the Fiction*
Since reality is much more complex than visions are, utopia becomes complicated as soon as we try to live in it. There is not only one utopia, but there are several ones – depending on different stakeholders. For example, what young people would call a perfect society might not be perfect for the elderly. In the last step (Figure 3) of our approach, we evaluated the enacted utopia with our participants. Therefore we made use of Anticipatory Ethnography [18] and developed it further. Other than Anticipatory Ethnography we did not evaluate utopia from a distanced point of view, but *from within* the fiction. After the enactments took place, we asked our participants to step in front of a camera and talk about the fictional technology – keeping their fictional character and perspective. Participants commented on what they experienced in this way: "My name is Marvin and I am the mayor of *Smalltown*, when I heard of *Hyperpipe* my first thought was 'this might be expensive but since our town is a little marooned we might benefit from it'. Since we installed it, people really enjoy sharing resources or even sending gifts with *Hyperpipe*. But I am increasingly getting the impression that people rarely leave their houses because of it. We really need to change our new habits before public life dies in *Smalltown*." This way we could collect differing viewpoints and an ambiguous perspective onto the former utopia.

## Conclusion
In this position paper we propose an approach that is (1) positive, (2) contextualized, (3) participatory and (4) evaluated from within the fiction. What started to be a utopian vision became a tangible and negotiable scenario with contradictory positive and negative emotions. Utopia developed a life of its own. Thus, we got insights into differing needs, hopes, expectations, and frustrations. We did not design fictional artifacts but fictional social interdependencies. The design of the fictional technology itself was not our focus.

In the workshop we would like to share our experiences with participatory, positive Design Fiction. Further, we would like to discuss the critical aspects of Design Fiction we identified and how far our approach is a constructive reaction to them.